\documentclass[twocolumn,showpacs,preprintnumbers,amsmath,amssymb,superscriptaddress]{revtex4}
\usepackage{epsfig,bm}
\usepackage{ulem}
\usepackage{amsmath,color}
\begin{document}
\title{Probing neutron-skin thickness with total reaction cross sections}
\author{W. Horiuchi}
\affiliation{Department of Physics, Hokkaido University, Sapporo 060-0810, Japan}
\author{Y. Suzuki}
\affiliation{Department of Physics, Niigata University, Niigata 950-2181, Japan}
\affiliation{RIKEN Nishina Center, Wako 351-0198, Japan}
\author{T. Inakura}
\affiliation{Department of Physics, Graduate School of Science, Chiba University, Chiba, 263-8522, Japan}

\begin{abstract}
We analyze total reaction cross sections, $\sigma_R$,
for exploring their sensitivity  
to the neutron-skin thickness of nuclei. We cover  
91 nuclei of O, Ne, Mg, Si, S, Ca, and Ni isotopes. 
The cross sections are calculated in the 
Glauber theory using the density distributions obtained 
with the Skyrme-Hartree-Fock method in 3-dimensional coordinate space.
Defining a reaction radius, $a_R=\sqrt{\sigma_R/\pi}$, to 
characterize the nuclear size and target (proton or $^{12}$C) dependence,  
we find an empirical formula for expressing $a_R$ with 
the point matter radius and the skin thickness, and 
assess two practical ways of determining the skin thickness 
from proton-nucleus $\sigma_R$ values measured at 
different energies  or from $\sigma_R$ values measured for different targets.
\end{abstract}
\maketitle

A systematic study of nuclear size properties 
tells us the saturation property of atomic 
nuclei. Recently the nuclear isovector size property, that is,  
neutron-skin thickness has attracted much interest.
The knowledge of skin thickness gives more insight into the 
properties of neutron-rich nuclei and neutron stars, and 
the equation of state (EOS) of asymmetric nuclear matter. 
For example, it is pointed out in 
Refs.~\cite{Chen10,RocaMaza11,Kortelainen13,Inakura13} that 
the skin thickness of finite nuclei constrains 
the symmetry energy and the slope parameter $L$
of the pure neutron matter EOS at the saturation density,
which is one of the key ingredients for the two-solar-mass 
neutron star problem~\cite{Demorest10}. 

The skin thickness has also been studied experimentally.
The parity-violating elastic electron scattering~\cite{PREX}
has been performed to determine the skin thickness of 
$^{208}$Pb, yielding 0.33$^{+0.16}_{-0.18}$ fm. 
Further measurement is planned to get more precise data
for $^{208}$Pb~\cite{PREXII}. 
Since the neutron radius is difficult to probe,
measurements of the skin thickness are still not as precise
as those of the proton radius, which is extracted from the charge 
distribution obtained by an electron scattering.

A hadronic probe is also useful to study the size properties of nuclei. 
Proton elastic scattering measurements at 295 MeV
have been done to probe the nuclear distributions 
of heavy stable targets, Sn~\cite{Terashima08} 
and Pb~\cite{Zenihiro10} isotopes.
Electric dipole response has been measured for $^{208}$Pb
using a $(p,p^{\prime})$ reaction 
and the skin thickness is evaluated as 0.156$^{+0.025}_{-0.021}$ fm 
by making use of the dipole polarizability 
of $^{208}$Pb~\cite{Tamii11}.
However, it is difficult to extend such measurements to unstable nuclei 
because of their short lifetimes.

Total reaction or interaction cross sections 
for unstable nuclei are more easily and accurately measured  
as long as they are produced sufficiently. 
Recent radioactive ion beam facilities allow us 
to measure precise total reaction cross sections for neutron-rich 
Ne and Mg isotopes on $^{12}$C target~\cite{Takechi10,Takechi.new,Kanungo11}.
The total reaction cross section 
on $^{12}$C target primarily probes the matter radius,
and therefore we need to know 
neutron or proton radii to determine the skin thickness.
The charge radii of unstable nuclei is made available 
by isotope shift measurements~\cite{Geithner08, Yordanov12, Krieger12}.
A combination of the deduced matter and proton radii 
from the different experiments gives us information on 
the skin thickness. 
The isotope shift measurement is, however, at present possible 
only for some limited cases, and moreover deducing the charge radius
from the measurement calls for extensive evaluations 
for various corrections. The measurement of charge-changing cross section 
may be an alternative to probe the proton radius, but 
that cross section does not always probe the proton 
radius directly, and thus one needs some model-dependent 
corrections to extract the proton radius~\cite{Yamaguchi11}. 

The purpose of this study is to 
discuss the possibility of 
using the total reaction cross 
sections to extract the skin thickness. 
Recalling the fact that the 
neutron-proton total cross section is larger than that of 
proton-proton below the incident energy of 550 MeV, we expect 
a proton target to probe more sensitively 
the neutron distribution in the tail region 
than $^{12}$C target~\cite{Ibrahim08}.
Based on the Glauber formalism~\cite{Glauber}, we systematically 
analyze the total reaction cross sections  for 
many nuclei with mass number $A=14-86$ of O, Ne, Mg, Si, S, Ca, and 
Ni isotopes. The wave functions of those nuclei 
are generated by the Skyrme-Hartree-Fock method 
on 3-dimensional coordinate space. 
The analysis of the cross sections enables us to propose 
possible ways to extract the skin thickness through the energy-
and target-dependence of the total reaction cross sections.

The total reaction cross section is calculated by
\begin{align}
\sigma_R=\int d\bm{b}\,(1-|\text{e}^{i\chi(\bm{b})}|^2),
\end{align}
where $\chi(\bm{b})$ is the phase-shift function for the elastic 
scattering of a projectile nucleus ($P$) and a target nucleus ($T$)
and the integration is done over the impact parameter $\bm{b}$.
The phase shift function is defined by a multiple integration 
with the ground state wave functions of the projectile and target
and its evaluation may be 
performed with a Monte Carlo technique as was done in Ref.~\cite{Varga02}. 
Here we use  
the so-called optical limit approximation (OLA), which requires  
only the one-body densities of the projectile 
and target, $\rho_P$ and $\rho_T$, respectively, 
\begin{align}
\text{e}^{i\chi(\bm{b})}
&=\exp\left[-\iint d\bm{r}^Pd\bm{r}^T \rho_P(\bm{r}^P)\rho_T(\bm{r}^T)\right.
\notag\\
&\times\Gamma_{NN}(\bm{s}^P-\bm{s}^T+\bm{b})\bigg],
\label{psf}
\end{align}
where $\bm{s}^P$ ($\bm{s}^T$) is the transverse component of 
$\bm{r}^P$ ($\bm{r}^T$) perpendicular to the beam direction, 
and $\Gamma_{NN}$ is the nucleon-nucleon ($NN$) profile function 
describing the $NN$ collision at  incident energy $E$. 
The profile function is different between proton-proton ($pp$) and 
proton-neutron ($pn$). The profile function 
for neutron-neutron is taken the same as $pp$. 
The integration in Eq.~(\ref{psf}) 
is carried out using the proton and neutron densities of the 
projectile and target. 

We use a usual parametrization for $\Gamma_{NN}$, 
\begin{align}
\Gamma_{NN}(\bm{b})=\frac{1-i\alpha_{NN}}{4\pi\beta_{NN}}
\sigma_{NN}^{\rm tot}\exp\left[-\frac{\bm{b}^2}{2\beta_{NN}}\right],
\label{profn.eq}
\end{align}
where $\alpha_{NN}$ is the ratio of the real to the imaginary part of the $NN$
scattering amplitude in the forward angle, $\beta_{NN}$ is
the slope parameter of the $NN$ elastic scattering differential cross section,
and $\sigma_{NN}^{\rm tot}$ is the total cross section of the $NN$ scattering. 
They are tabulated in Ref.~\cite{Ibrahim08} for a wide range of energy.
Though it misses some higher-order terms of $\Gamma_{NN}$,  
the OLA describes the proton-nucleus 
scattering satisfactorily~\cite{Varga02}. However, 
the OLA for nucleus-nucleus collisions  
tends to predict larger cross sections than measurement~\cite{Horiuchi07}, 
and we employ another expression called 
the nucleon target formalism in the Glauber theory (NTG)~\cite{Ibrahim00}.
The NTG requires the same inputs as the OLA
and reproduces  $\sigma_R$ fairly well. 
For example, $\sigma_R$ of $^{12}$C+$^{12}$C calculated with NTG  
are improved very much in a wide energy range~\cite{Horiuchi07}.
The power of NTG is also confirmed in 
applications to $^{22}$C~\cite{Horiuchi06}, 
Oxygen isotopes~\cite{Ibrahim09} as well as 
light neutron-rich nuclei~\cite{Horiuchi10,Horiuchi12}.
In this paper we employ the OLA for proton target
and the NTG  for $^{12}$C target.

For projectiles with large $Z$, the Coulomb force contributes 
to $\sigma_R$ via Coulomb breakup. Its effect may 
be taken into account by, e.g., the Coulomb corrected 
eikonal approximation~\cite{Margeron03,Ibrahim04,Capel08}. 
Since $Z$ involved in the present calculation is not very large, 
the Coulomb interaction between the projectile and target is 
ignored for the sake of simplicity. 

We perform the Skyrme-Hartree-Fock (HF) calculation for the density
distribution of a variety of projectiles.
The ground state is obtained by minimizing the energy density
functional~\cite{VB72}.
Every single-particle wave function
is represented in the 3-dimensional 
grid points with the mesh size of 0.8\,fm.
All the grid points inside the sphere of radius of 20\,fm are
adopted in the model space.
The ground state is constructed by the imaginary-time method~\cite{DFKW80}.
The angle-averaged intrinsic one-body densities are
used as the one-body densities required for the Glauber calculation.
For more detail, see Ref.~\cite{Horiuchi12}.

We consider light to medium mass even-even nuclei with $Z=8-16,\, 20$, 
and 28, covering both proton and neutron-rich region, that is, 
$^{14-24}$O, $^{18-34}$Ne, $^{20-40}$Mg, $^{22-46}$Si, $^{26-50}$S, 
$^{34-70}$Ca, and $^{48-86}$Ni.
Two Skyrme parameter sets,
SkM* \cite{bartel82} and SLy4 \cite{Chan97}, are employed. 
The SkM* functional is known to well account for the 
properties of the nuclear deformation,
while the SLy4 is superior to SkM* in reproducing 
the properties of neutron-rich nuclei. 
In Ref.~\cite{Horiuchi12} we study light
neutron-rich nuclei with even proton numbers $Z=8-16$ and 
discuss the deformation effects on $\sigma_R$ for $^{12}$C target. 
The two interactions give a significant difference in
$\sigma_R$ due to the nuclear deformation~\cite{Horiuchi12}.

As is well known, $\sigma_R$ carries information on the nuclear size. 
It is convenient to define a `reaction radius', $a_R$, 
of the nucleus-nucleus collision at incident energy $E$ by 
\begin{align}
a_R (N,Z,E,T)=\sqrt{\sigma_R (N,Z,E,T)/\pi},
\end{align}
where $N$ and $Z$ are the neutron and proton numbers of the projectile, 
and $T$ stands for the target, either proton or  $^{12}$C. In what follows 
we omit $T$ in most cases and mean by $E$  the projectile's 
incident energy per nucleon. 
The reaction radius depends on $E$  
through the energy dependence of the profile function~(\ref{profn.eq}).
The $pn$ and $pp$ total cross sections, 
$\sigma_{pn}^{\rm tot}$ and $\sigma_{pp}^{\rm tot}$, 
behave differently as follows~\cite{PDG}. 
At $E=100$ MeV, $\sigma_{pn}^{\rm tot}$ is about 
three times larger than $\sigma_{pp}^{\rm tot}$. 
The difference between them gets smaller as the energy increases,  
and vanishes at about 550 MeV, and beyond 800 MeV 
$\sigma_{pp}^{\rm tot}$ exceeds $\sigma_{pn}^{\rm tot}$. 
The above energy-dependence of the basic 
inputs of $NN$ data is best reflected in $\sigma_R$ for 
proton target, whereas it is averaged out in the case of $^{12}$C 
target. Thus we expect the proton target to be advantageous for 
probing the neutron-skin part.

\begin{figure}[th]
\begin{center}
\epsfig{file=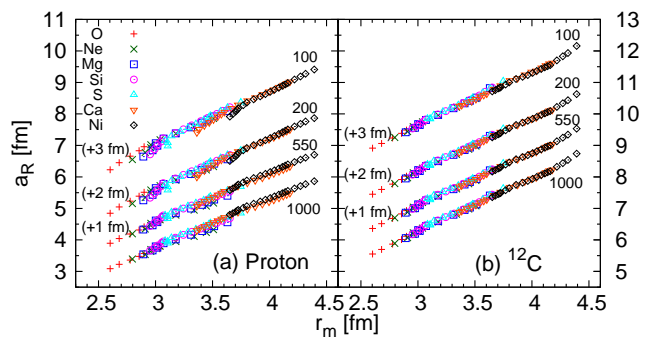,scale=0.82}
\caption{(Color online). 
Reaction radii vs. point matter rms radii for 
O, Ne, Mg, Si, S, Ca, and Ni isotopes 
on (a) proton and (b) $^{12}$C targets at incident energies of 
100, 200, 550, and 1000 MeV. The SkM* interaction is used.
The reaction radii of 550, 200, and 100 MeV
are added by 1, 2, and 3\, fm, respectively, for the sake of illustration.}
\label{eradrm.fig}
\end{center}
\end{figure}

We define the matter radius $r_m(N,Z)$ and the skin thickness $\delta(N,Z)$ 
by 
\begin{align}
&r_m(N,Z)=\sqrt{{\textstyle{\frac{Z}{A}}}r_p^2(N,Z)+{\textstyle{\frac{N}{A}}}r_n^2(N,Z)},\notag \\
&\delta(N,Z)=r_n(N,Z)-r_p(N,Z),
\end{align}
where $r_p(N,Z)$ and $r_n(N, Z)$ are point proton and point 
neutron root-mean-square (rms) radii, respectively. 
Figure~\ref{eradrm.fig} displays 
the reaction radii of the 91 nuclei as a function of $r_m (N,Z)$. 
As expected, $a_R$ shows a linear dependence on $r_m (N,Z)$ at 
any incident energies, which is the basis of 
extracting the matter radius from $\sigma_R$.  
A closer look at the proton case shows, however, somewhat 
scattered distributions along the straight lines. 
This indicates that the reaction radii for proton target 
scatter depending on 
the detail of the neutron and proton density profiles even though 
$r_m (N,Z)$ values are the same, 
namely the reaction radii for the proton target 
carry some information on the skin thickness. The same conclusion is 
drawn with SLy4 as well. 

\begin{figure}[th]
\begin{center}
\epsfig{file=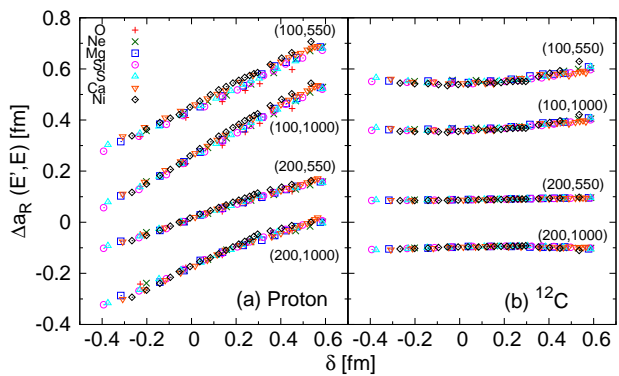,scale=0.78}
\caption{(Color online). Difference of the reaction radii 
at two incident energies $(E^\prime, E)$ in MeV 
as a function of the skin thickness
for O, Ne, Mg, Si, S, Ca, and Ni isotopes 
on (a) proton and (b) $^{12}$C targets.
The SkM* interaction is used.}
\label{eradpCskinSkMs.fig}
\end{center}
\end{figure}

To substantiate the above statement on probing the neutron-proton 
density profiles, we consider the difference of the reaction radii
at two incident energies
\begin{align}
\Delta a_R (N,Z,E^\prime, E)=a_R(N,Z,E^\prime)-a_R(N,Z,E).
\end{align}
Figure~\ref{eradpCskinSkMs.fig} plots $\Delta a_R$  
as a function of $\delta(N,Z)$. Four different sets 
of energies are chosen. 
The isotope dependence of $\Delta a_R$ is weak for both targets,  
and in each set $\Delta a_R$ for all the nuclei 
tend to approximately follow a straight line. 
For proton target $\Delta a_R$ vs. $\delta(N,Z)$ diagram exhibits strong 
correlation as revealed by nonzero slope, that is, $\Delta a_R$ can be 
a quantity sensitive to the skin thickness.  
The steepest slope among the four sets is obtained 
for  $(E^\prime, E)=(100,1000)$ MeV, which is 
easily understood from 
the energy dependence of $\sigma_{pn}^{\rm tot}$ and $\sigma_{pp}^{\rm tot}$. 
$\Delta a_R$ values for $^{12}$C target show flat behavior, or the 
slope is almost zero, which is a consequence of the fact that 
the $pn$ and $pp$ profile functions are averaged because of the equal 
numbers of protons and neutrons in $^{12}$C. 
Thus the $^{12}$C target is never sensitive to 
the skin thickness.    

\begin{figure}[bh]
\begin{center}
\epsfig{file=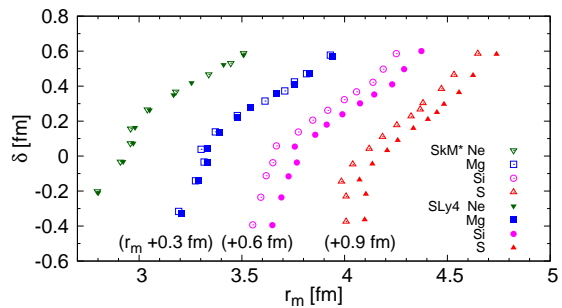,scale=0.75}
\caption{(Color online). 
Comparison of the skin thickness and point matter rms radius of 
Ne, Mg, Si, and S isotopes calculated with 
the SkM* and SLy4 interactions. The radii of Mg, Si, 
and S isotopes are added by 0.3, 0.6, and 0.9\, fm, respectively, 
for the sake of illustration.}
\label{rmsskin.fig}
\end{center}
\end{figure}

We confirm that the above observation holds the case in the SLy4 case
as well, namely $\Delta a_R$ vs. $\delta(N,Z)$ diagram obtained 
with the SLy4 interaction is almost the same as Fig.~\ref{eradpCskinSkMs.fig} 
obtained with SkM*. It appears that this is never trivial because  
the SkM* and SLy4 interactions predict different density profiles as 
shown in Fig.~\ref{rmsskin.fig}, which compares  
$r_m(N,Z)$ and $\delta(N,Z)$ values of Ne, Mg, Si, and S isotopes 
between the two sets.
As discussed in Ref.~\cite{Horiuchi12}, the matter radii
of the Ne and Mg isotopes are correlated with their deformations that 
strongly depend on the Skyrme interactions. 
In fact we see the interaction dependence of $\sigma_R$ or $a_R$ 
in particular in $Z=10-16$ region  as shown 
in Fig.~\ref{rmsskin.fig},  
whereas very similar results are obtained for O, Ca, Ni isotopes. 
Despite this difference the two Skyrme interactions 
give virtually the same $\Delta a_R$ vs. $\delta(N,Z)$ diagram.  
Thus $\Delta a_R$ is not sensitive 
to the nuclear shape and the density distribution either but 
is sensitive to the skin thickness. 

\begin{figure}[tb]
\begin{center}
\epsfig{file=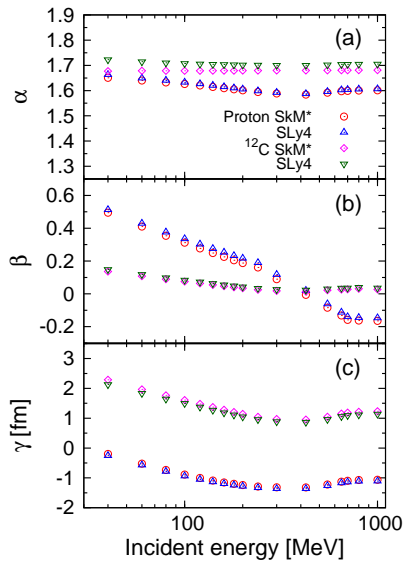,scale=0.85}
\caption{(Color online). Coefficients of reaction radius,  (a) $\alpha(E)$,
(b) $\beta(E)$, and (c) $\gamma(E)$ as a function of projectile's 
incident energy per nucleon. 
The SkM* and SLy4 interactions are used.}
\label{eradpCfit.fig}
\end{center}
\end{figure}

The approximate linear dependence of both $a_R$ on 
$r_m(N,Z)$ and $\Delta a_R$ on $\delta(N,Z)$ suggests the following ansatz 
\begin{align}
a_R(N,Z,E)=\alpha(E)r_m(N,Z)+\beta(E)\delta(N,Z)+\gamma(E),
\label{fit.eq}
\end{align}
where $\alpha(E)$, $\beta(E)$, and $\gamma (E)$ depend on 
the target as well. Those coefficients are 
determined by minimizing the mean square deviation, 
$\chi^2(E)=\sum_{N,Z}[a_R^{\rm HF}(N,Z,E)-a_{R}^{\rm Fit}(N,Z,E)]^2/{\cal N}$,
where $a_R^{\rm HF}$ and $a_R^{\rm Fit}$ are $a_R$ values 
calculated with the HF densities and Eq.~(\ref{fit.eq}), respectively, 
and $\mathcal{N}=91$ is the number of data points. 
We cover 17 incident energies from 40 to 1000\,MeV. 
It may be questionable to apply the Glauber theory to such low energy 
as $E=40$\, MeV for proton target, but we include it  for the sake of 
convenience. 
The square root of $\sum_i \chi^2(E_i)/17$ for proton target is found to be 
0.052 fm for SkM* and 0.049 for SLy4, 
which is approximately 1\% of the mean  $a_R$ value, 
4.68 and 4.70 fm, of all the nuclei at 17 energy points, respectively. 
As displayed in Fig.~\ref{eradpCfit.fig}, the resulting coefficients  
show `universality', that is, very weak dependence on the Skyrme 
interactions, especially in the proton case. 
It should be stressed that Eq.~(\ref{fit.eq}) is valid 
for all the nuclei considered in this paper. 
We tried another ansatz by replacing $\delta(N,Z)$ with 
$\delta(N,Z)/r_m(N,Z)$, but the result was worse. 

Both $\alpha(E)$ and $\gamma(E)$ weakly depend on $E$ 
above 100\, MeV. The term with $\alpha(E)$ is a major contributor 
to $a_R$ among the three terms, and its weak 
energy dependence of less than few \% for both proton and $^{12}$C targets 
explains the linearity of $a_R$ on $r_m(N,Z)$ with almost equal slopes 
as displayed in Fig.~\ref{eradrm.fig}.
The energy dependence of $\beta(E)$ is strong for proton target. 
This is responsible for the scattered distributions 
of the reaction radii for the proton case as noted in Fig.~\ref{eradrm.fig}. 
It is easy to understand that the linear dependence of $\Delta a_R$ on 
$\delta(N,Z)$ shows up in the proton case but almost disappears in $^{12}$C 
target as the main $\alpha(E)$ terms are canceled out. 
In this way the proton target probes the skin thickness sensitively 
but the $^{12}$C target does not. 
The energy dependence of $\gamma(E)$ is similar 
between  proton and $^{12}$C targets. The latter is obtained by 
adding about 2.4 fm to $\gamma(E)$ 
of the proton target, which is understandable considering that 
the reaction radius of $^{12}$C target includes 
not only the radius of the projectile but also that of $^{12}$C.

We have determined $\alpha(E)$, $\beta(E)$, and $\gamma (E)$
using the results on the 91 nuclei obtained with the two Skyrme interactions. 
It will be interesting to further test the universal relation between $a_R$ 
and $r_m(N,Z)$ as well as  $\delta(N,Z)$ 
in heavier nuclei using different Skyrme interactions.

\begin{table}[ht]
\caption{$p+^{40}$Si total reaction cross sections 
in mb at incident energy $E$ in MeV calculated 
with different Fermi distributions. Each of the distributions 
is specified by the 
proton and neutron diffuseness parameters in fm, $a_p$ and $a_n$, and 
the proton and neutron radius parameters are set to reproduce 
the matter radius (3.46 fm) and skin thickness (0.37 fm) 
of $^{40}$Si obtained with the SkM* HF calculation. See Fig.~\ref{fermi.fig}.} 
\begin{center}
\begin{tabular}{cccccccccccc}
\hline\hline
$(a_p,a_n)$\textbackslash$E$&&100&120&140&160&200&300&425&550&800&1000\\
\hline
(0.5, 0.5) &&742&700&667&642&607&564&550&575&615&619\\    
(0.6, 0.6) &&752&706&672&646&609&565&551&577&619&622\\ 
(0.7, 0.7) &&759&711&675&648&609&564&549&576&621&623\\
(0.5, 0.7) &&756&708&673&646&608&563&549&575&617&620\\
HF(SkM*)   &&747&703&670&644&608&565&551&576&617&620\\
\hline\hline
\end{tabular}
\end{center}
\label{fermi.tab}
\end{table}

\begin{table}[ht]
\caption{Same as Table~\ref{fermi.tab} but on $^{12}$C target
in units of b.}
\begin{center}
\begin{tabular}{ccccccccccc}
\hline\hline
$(a_p,a_n)$\textbackslash$E$&&100&140&160&200&300&425&550&800&1000\\
\hline
(0.5, 0.5) &&1.73&1.62&1.58&1.52&1.45&1.44&1.48&1.55&1.56\\    
(0.6, 0.6) &&1.78&1.66&1.62&1.56&1.48&1.47&1.51&1.59&1.60\\ 
(0.7, 0.7) &&1.84&1.71&1.67&1.60&1.52&1.51&1.55&1.64&1.65\\
(0.5, 0.7) &&1.80&1.68&1.64&1.57&1.49&1.48&1.53&1.61&1.62\\
HF(SkM*)   &&1.74&1.63&1.59&1.53&1.45&1.44&1.49&1.56&1.57\\
\hline\hline
\end{tabular}
\end{center}
\label{fermiC.tab}
\end{table}

\begin{figure}[ht]
\begin{center}
\epsfig{file=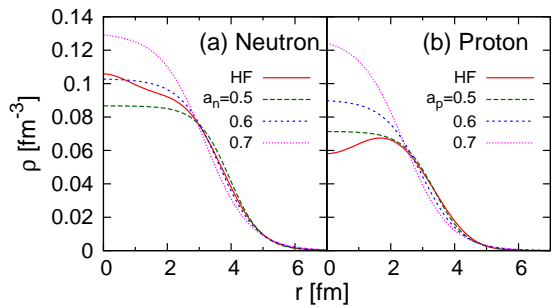,scale=0.9}
\caption{(Color online). Matter, neutron, and proton densities 
of $^{40}$Si of Fermi distributions  
with different sets of diffuseness parameter.
The HF densities with SkM* are also plotted for comparison.}
\label{fermi.fig}
\end{center}
\end{figure}

Because Eq.~(\ref{fit.eq}) is found to be valid to good approximation, 
we can make use of it to know the skin thickness.
Given that two $\sigma_R$ or $a_R$ values for a proton target 
are accurately measured at different energies, 
we can extract $r_m(N, Z)$ and $\delta(N, Z)$ values as follows. 
A crude estimate of those unknowns 
is to solve a simultaneous linear equation based on Eq.~(\ref{fit.eq}). 
Since precise values of $\alpha(E)$, $\beta(E)$, and $\gamma(E)$ 
may  actually be unknown, another way without recourse to them 
is preferable. 
We assume some model density 
distributions for protons and neutrons and calculate 
$\sigma_R$ values.  If those cross sections agree with the measured cross 
sections simultaneously, $r_m(N, Z)$ and $\delta(N, Z)$ are 
calculated from the assumed distributions. Otherwise 
other density distributions are tested 
until the model distributions reproduce the measured cross sections.
Here the model distribution may be, for instance, Fermi-type. 
This assertion is based on the fact that 
$\sigma_R$ is not sensitive to the detail of the density distribution 
but is determined by $r_m(N, Z)$ and $\delta(N, Z)$. Conversely, 
two different density distributions predict 
virtually the same $\sigma_R$ value in so far as they 
give same $r_m(N, Z)$ and $\delta(N, Z)$ values. To corroborate the 
statement, we compare in Table~\ref{fermi.tab} 
$\sigma_R$ values of $p+^{40}$Si calculated using Fermi  
distributions with different diffuseness parameters.
The proton and neutron radius parameters of each distribution is set 
to reproduce the $r_m(N, Z)$ and $\delta(N, Z)$ values of the SkM* density. 
The tail parts of the densities 
are crucially important to determine $\sigma_R$. 
Though the shapes of the distributions are different
as shown in Fig.~\ref{fermi.fig},
all the densities give $\sigma_R$ very close 
to the HF cross section at all the energies. Even at 100\, MeV 
the deviation from the HF result falls within at most 1.6 \%,  
confirming our statement. 
For a practical measurement, the two energies are to be chosen 
from low and high energy regions, e.g. 
$E\lesssim 200$ and $E\gtrsim 550$\, MeV to make use 
of the sensitivity to the skin thickness as shown in
the energy dependence of $\beta(E)$
and the difference of the two cross sections has to be large enough 
to be distinguished beyond experimental uncertainties.

Another possible way is to measure
$\sigma_R$ values on proton and $^{12}$C targets.
Assuming some model densities for protons and neutrons,
$r_m(N, Z)$ and $\delta(N, Z)$ can also be determined 
by reproducing the $\sigma_R$ simultaneously.
The incident energy for $^{12}$C target may be chosen arbitrarily because
the $a_R$ on $^{12}$C target is only sensitive to $r_m(N,Z)$
as shown in Fig.~\ref{eradpCfit.fig}.
For proton target, it is advantageous to choose low energy, say $E\lesssim 200$ 
for maximizing the sensitivity to $\delta(N,Z)$ as much as possible.
Table~\ref{fermiC.tab} lists $\sigma_R$ on $^{12}$C target obtained with  
the different Fermi distributions.
The deviation from the HF result is about 5 \% for all $E$, 
which is not as small as that of proton target. 
Because of this we must 
admit that constraining $r_m(N,Z)$ through $\sigma_R$ for $^{12}$C target 
contains some uncertainty.
In Ref.~\cite{Kanungo11}, $\sigma_R$ for proton and $^{12}$C targets
are measured for neutron-rich Mg isotopes at 900 MeV, and 
$r_m(N,Z)$ of $^{32-35}$Mg isotopes are determined within approximately 
5\% in the analysis with the Fermi density distributions.  
Unfortunately $\sigma_R$ on a 'proton' target did not set 
a constraint on $\delta(N, Z)$ because the proton data contained  
large uncertainty and the profile function at 900\, MeV is insensitive  
to the neutron-skin thickness. It is necessary to reduce 
the experimental uncertainty of $\sigma_R$ on proton target
for a precise determination of $r_m(N,Z)$ and $\delta(N,Z)$.

In summary, to explore a sensitive probe to the skin thickness, 
we have made a systematic analysis of total reaction
cross sections, $\sigma_R$, on proton and $^{12}$C targets in 
the Glauber model. The Skyrme-Hartree-Fock
method is applied to generate the  densities of  91 even-even nuclei with $A=14-86$ from 
$Z=8-16$, 20, and 28 elements. Two different interactions, SkM* and SLy4, are 
employed to test the nuclear size properties. 

We find a universal expression that linearly relates  
the reaction radius, $a_R=\sqrt{\sigma_R/\pi}$, to 
the point matter radius $r_m(N,Z)$ and skin thickness $\delta(N,Z)$. 
The proportional coefficients are determined as a function of the 
incident energy by analyzing $\sigma_R$ calculated 
from the Hartree-Fock densities. Among others, the coefficient 
proportional to $\delta(N,Z)$ in the case of proton target exhibits remarkable 
energy dependence. This sensitivity of $a_R$ or $\sigma_R$ 
to the skin thickness has enabled us to assess a practical way of 
determining both $\delta(N,Z)$ and $r_m(N,Z)$ from the measurements 
of proton-nucleus reaction cross sections at two different energies 
 or from a combination of two $\sigma_R$ values measured 
on both $^{12}$C and proton targets. 

The work was in part supported by JSPS KAKENHI Grant Numbers (24540261 and
25800121).

\end{document}